\documentclass[conference,final]{IEEEtran}
\usepackage[english]{babel}
\usepackage{url}
\usepackage{cite}
\usepackage{amsmath,amssymb,amsfonts}
\usepackage{graphicx}
\usepackage{caption}
\usepackage{textcomp}
\usepackage{xcolor}
\usepackage{verbatim}
\usepackage{xspace}
\usepackage{algorithm,algpseudocode}
\usepackage{cite}
\usepackage{amsmath,amssymb,amsfonts}
\usepackage{graphicx}
\usepackage{textcomp}

\newcommand{\singlelist}{\texttt{SingleList}{}\xspace}
\newcommand{\nocachelist}{\texttt{NoCacheList}{}\xspace}
\newcommand{\linkedlist}{\texttt{LinkedList}{}\xspace}
\newcommand{\arraylist}{\texttt{ArrayList}{}\xspace}
\newcommand{\arrayring}{\texttt{ArrayRing}{}\xspace}
\newcommand{\arrayblock}{\texttt{ArrayBlock}{}\xspace}
\newcommand{\collection}{\texttt{Collection}{}\xspace}

\newcommand{\java}{\texttt{Java}{}\xspace}
\newcommand{\cpp}{\texttt{C++}{}\xspace}
\newcommand{\lisaac}{\texttt{Lisaac}{}\xspace}

\newcommand{\data}[1]{{\tt{}data}\,\#{\tt{}#1}}

\newcommand{\datastructfig}[2]{%
\begin{figure*}[!ht]%
Memory Layout of {\texttt{#1}}\vspace{15pt}\\%
\includegraphics[width=0.81\textwidth]{#1.eps}%
\caption{#2}%
\label{#1}%
\end{figure*}%
}

\newcommand{\benchmarkfig}[2]{%
\begin{figure}%
\begin{center}%
\includegraphics[width=8.75cm]{#1.eps}%
\end{center}%
\caption{#2}%
\label{#1}%
\end{figure}%
}

\def\BibTeX{{\rm B\kern-.05em{\sc i\kern-.025em b}\kern-.08em
    T\kern-.1667em\lower.7ex\hbox{E}\kern-.125emX}}
\makeatletter
 \let\old@ps@headings\ps@headings
 \let\old@ps@IEEEtitlepagestyle\ps@IEEEtitlepagestyle
 \def\confheader#1{%
 \def\ps@IEEEtitlepagestyle{%
 \old@ps@IEEEtitlepagestyle%
 \def\@oddhead{\strut\hfill#1\hfill\strut}%
 \def\@evenhead{\strut\hfill#1\hfill\strut}%
 }%
 \ps@headings%
 }
 \makeatother

\confheader{%
 International Journal of Future Computer and Communication
 }

 \usepackage[pscoord]{eso-pic}
\newcommand{\placetextbox}[3]{
 \setbox0=\hbox{#3}
 \AddToShipoutPictureFG*{ \put(\LenToUnit{#1\paperwidth},\LenToUnit{#2\paperheight}){\vtop{{\null}\makebox[0pt][c]{#3}}}
 }
 }
 \placetextbox{.23}{0.055}{\small{979-8-3503-8166-5/24/\$31.00~\copyright 2024 IEEE}}

\def\BibTeX{{\rm B\kern-.05em{\sc i\kern-.025em b}\kern-.08em
    T\kern-.1667em\lower.7ex\hbox{E}\kern-.125emX}}

\begin{document}

\title{RIP Linked List\\
  {\Large{}Empirical Study to Discourage You from Using Linked Lists Any Further}}

\author{\IEEEauthorblockN{Benoît Sonntag}
\IEEEauthorblockA{\textit{Université de Strasbourg} \\
Strasbourg, France \\
Benoit.Sonntag@lisaac.org\\
0009-0000-2806-1970}
\and
\IEEEauthorblockN{Dominique Colnet}
\IEEEauthorblockA{\textit{Université de Lorraine - LORIA} \\
Nancy, France \\
Dominique.Colnet@loria.fr\\
0009-0006-7368-2235}
}

\maketitle

\begin{abstract}
Linked lists have long served as a valuable teaching tool in programming.
However, the question arises: Are they truly practical for everyday program use?
In most cases, it appears that array-based data structures offer distinct advantages, particularly in terms of memory efficiency and,
more importantly, execution speed.
While it's relatively straightforward to calculate the complexity of operations, gauging actual execution efficiency remains a challenge.
This paper addresses this question by introducing a new benchmark.
Our study compares various linked list implementations with several array-based alternatives.
We also demonstrate the ease of incorporating memory caching for linked lists, enhancing their performance.
Additionally, we introduce a new array-based data structure designed to excel in a wide range of operations.
\end{abstract}


\datastructfig{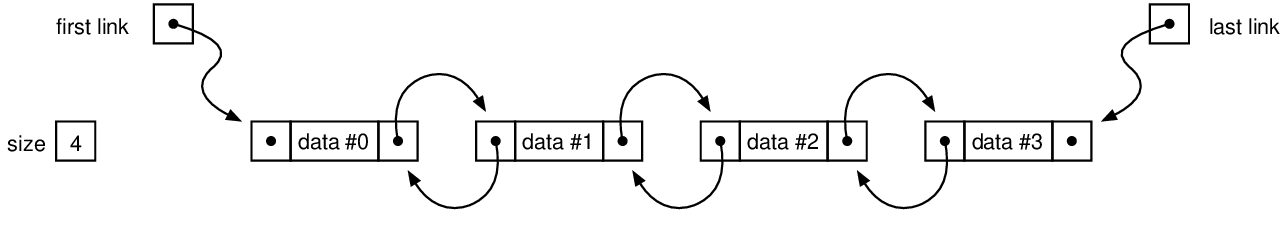}{
  A doubly linked list with a memory cache for {\tt{}size}, containing 4 data items.
  The head pointer is stored in {\tt{}first link}, and the tail pointer in {\tt{}last link}.
}

\datastructfig{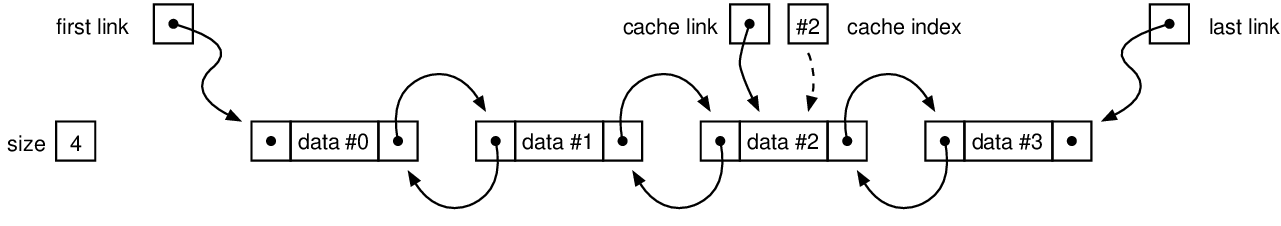}{
  Same layout as the one of Fig. \ref{NoCacheList}, but with an extra cache of the last visited index ({\tt{}cache\,link} and {\tt{}cache\,index}).
}

\datastructfig{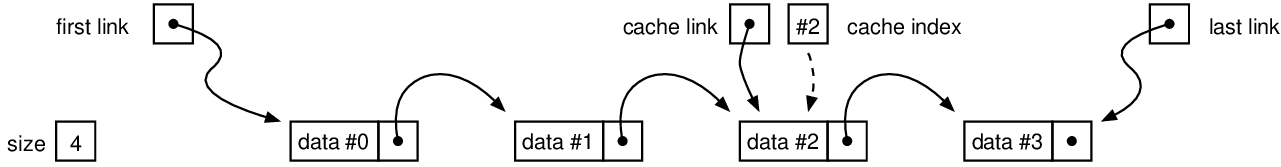}{
  One-way linked list with memory cache for the {\tt{}size} and the last visited index ({\tt{}cache\,link} and {\tt{}cache\,index}).
}

\section{Introduction}\label{intro}

This article is a feedback and practical analysis of list data structures.
After many years of programming practice, we realized that we never use a linked list anymore.
Is this famous list structure implementation that we all studied at one time or another during our studies really useful ?

The theoretical advantages of a linked list are however numerous and attractive:
\begin{enumerate}
\item
  It allows a constant incremental allocation of the memory.
  Indeed, the addition of an element is equivalent to the allocation of a single cell in the list.
\item
  There are never any memory moves of cells during the life of the linked list.
\item
  Knowing the location of the insertion or removal of an element, the operation requires a constant number of instructions.
\end{enumerate}
As for the drawbacks, it is the necessity of a partial and sequential path from cell to cell to reach an ith element that degrades performance.
Many more complex implementations are possible to partially compensate for this shortcoming.
We study two of them here: the presence of a backward chaining ("doubly linked" list), and the index cache management.
When computers did not have much RAM and the speed of moving from one memory area to another was still critical, advantages 1 and 2 made the linked list profitable.
In this paper, we want to know if there are still situations where the linked list is an advantageous and efficient implementation.

Eager to have a concrete and recent study of the structures in list and in search of the best strategy,
the Bjarne Stroustrup's benchmark\footnote{Bjarne Stroustrup's keynote in GoingNative 2012: Why you should avoid Linked List.
https://www.youtube.com/watch?v=YQs6IC-vgmo
} seems to provide elements of an answer.
Here, we propose to pursue the study and analyze the behavior of different list implementations using B. Stroustrup's benchmark.
In this study, we introduce a new implementation called {\arrayblock}, with the claim to have a relatively advantageous behavior in all real-life usage scenarios.
We also propose another benchmark, we called {\tt{}Fairbench}, which seems more relevant to test the efficiency and the behavior of linked lists in conditions
closer to a realistic usage.

Note also that there is a lot of educational material about linked lists and/or the use of arrays,
there are also some youtube vidéos \cite{youtubeBStroustrup2012,Computerphile,CalebCurry},
also some web articles \cite{BaptisteWicht1,BaptisteWicht2,DatHoang,Johnnylab}, 
but we found no research publication directly related to the main topic of this article.

\section{Linked-lists Representations}

We explore three linked list implementation options:
{\nocachelist} (Fig. \ref{NoCacheList}), {\linkedlist} (Fig. \ref{LinkedList}) and {\singlelist} (Fig. \ref{SingleList}).

Fig. \ref{NoCacheList} shows the memory representation of {\nocachelist}.
It is a doubly linked list that stores both the head and tail pointers.
To avoid having to traverse the list to find out its length, a memory cache, called {\tt{}size}, is used to store this information.
In the example of Fig. \ref{NoCacheList}, the list holds 4 pieces of data.
Note that this representation corresponds exactly to the {\tt{}LinkedList} class of the \java{} standard library\footnote{
The \java{} 8 (or later version of the language) is exactely like {\nocachelist}.
At the time we are writing this article, there is still no other cache than the {\tt{}size} cache.
}.
As in \java{}, index 0 allows access to the first element (\data{0}), the second is at index 1 (\data{1}), and so on.
Of course, if an element is added or removed, the {\tt{}size} attribute must be updated accordingly.

The memory cache technique can also be used to store the location corresponding to the last access made in the list (see {\linkedlist} on Fig. \ref{LinkedList}).
The two variables, {\tt{}cache\,index} and {\tt{}cache link} store the user index and the pointer to the last accessed link respectively.
In the example in Fig. \ref{LinkedList}, if the user wants to access \data{2}, the fastest way is to go through the index cache.
Thanks to the double linking and the index cache, sequential runs, from left to right or also from right to left, are in constant time for any list size.

The third representation considered, {\singlelist} on the Fig. \ref{SingleList}, corresponds to a single-linked list and,
as in the previous case, has a memory cache for both the index and the size.
Obviously, because of its single chaining, a {\singlelist} will only be efficient when traversing from the left to the right.
With the three previous representations, {\nocachelist}, {\linkedlist} and {\singlelist}, we cover the different possibilities for linked lists in a relatively exhaustive way.

One of the major drawbacks of linked lists is the amount of memory used by pointers.
Even though the memory addresses of today's machines are limited to 48 bits, due to re-alignment problems, each pointer currently costs 64 bits.
Thus, for a list of integers or pointers, the memory space of a list of $N$ elements in case of double linking is $N\times{}3\times{}64$ bits, that is $N\times{}24$ bytes.

\section{Array-based representations}
Using contiguous memory areas (i.e., native arrays) saves memory space.
For array-based representations, we have chosen two standard, well-known forms: {\arraylist} (Fig. \ref{ArrayList}) and {\arrayring} (Fig. \ref{ArrayRing}).
Finally, the third representation is the implementation we call {\arrayblock} (Fig. \ref{ArrayBlock}).

\datastructfig{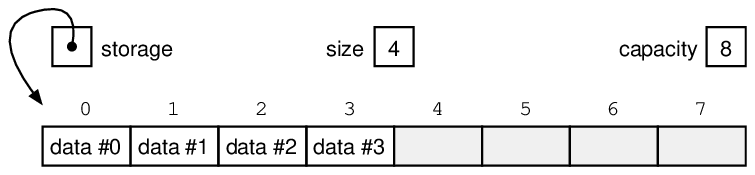}{
  Used area on the left and supply area on the right.
  Same indexing in the native array and in the user interface.
}

\datastructfig{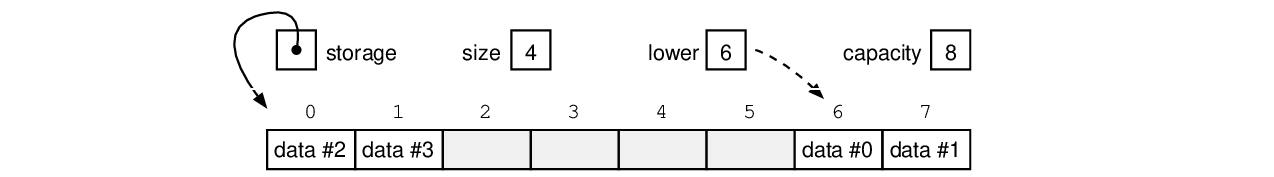}{
  The storage area is used in a circular fashion, from left to right.
  The variable {\tt{}lower} is used to locate the internal index of \data{0}.
}

\datastructfig{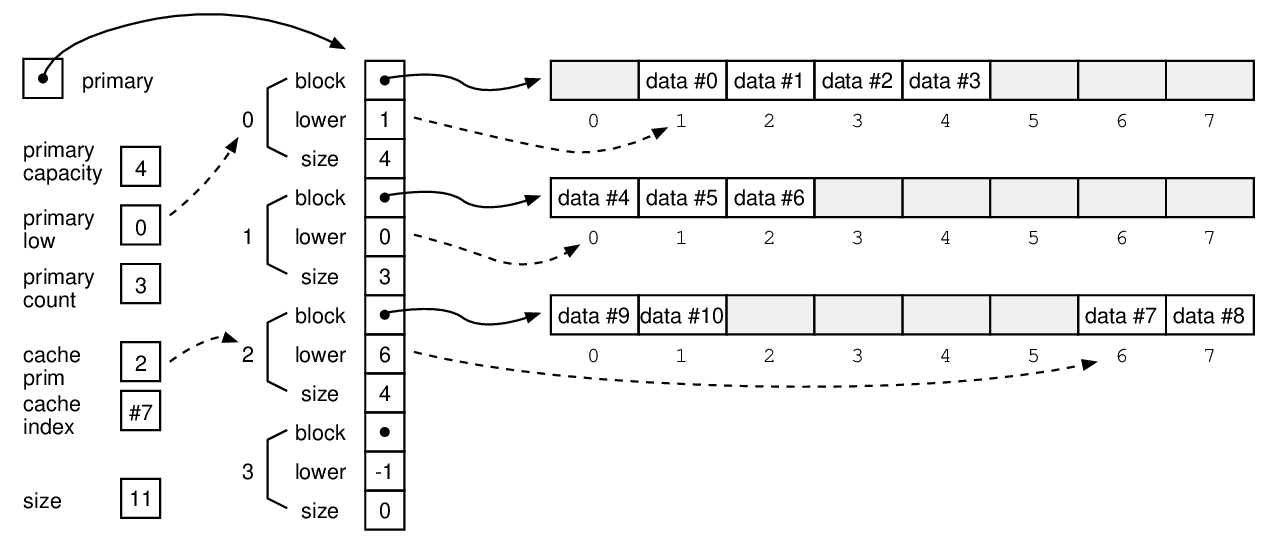}{
  A resizable primary table and storage fixed-size areas in power of 2.  
  Circular management of all the tables.
}

The {\arraylist} representation shown in the Fig. \ref{ArrayList} is quite common in programming languages libraries.
This representation has exactly the same name in the \java{} library.
In \cpp{} this data structure is also known as {\tt{}std::vector}.
The principle of this data structure is to provide a storage area that is at least equal to, and often larger than what is needed,
to avoid having to constantly adjust the size of the corresponding memory array \cite{colnet2015}.
In the example of the Fig. \ref{ArrayList}, the storage memory block consists of 8 slots, 4 of which are used and 4 are in reserve.
The variable {\tt{}storage} holds the pointer to the storage area and the variable {\tt{}capacity} holds the allocation size of the storage area.
The variable {\tt{}size} stores the fact that only 4 slots are used.
From the user's point of view, in order to comply with the same access interface as for the lists,
the {\tt{}4} stored datas are accessible via the index interval {\tt{}[0,size-1]}.
This representation is very simple because the access to the storage area is done without having to modify the index given by the user.
This array representation is particularly well-suited for adding/deleting in queue.
For example, deleting the last data item is simply a decrement of {\tt{}size}.
In the case of adding at the last position, if there are available slots in reserve, the operation is also trivial.
Obviously for an insert or an addition at the beginning, the operations become more complicated.
For example, to insert at the first position, all the elements must be shifted one place to the right in order to make room for the new element at the index {\tt{}0}.

Althought the memory capacity is twice the number of elements, the memory used is $N\times{}2\times{}64$ bits, that is $N\times{}16$ bytes.
Thus with a reserve area of the same size as the used area, the memory consumption remains reasonable compared to the space taken up by a doubly linked list (i.e. $N\times{}24$ bytes).

The Fig. \ref{ArrayRing} gives an example of the {\arrayring} representation which allows to solve the problem of the addition in the first position quite simply.
The principle is to use the storage area in a circular way.
To do this, we add a variable {\tt{}lower} that allows us to know where the data that the user accesses with index {\tt{}0} is located.
In the storage area, starting from this point, the data are stored from left to right, and, when we reach the end of the storage area, we start again from the beginning.
The math that gets you from the user index to the storage area index is just an addition with {\tt{}lower}.
Whether it is a leading or trailing addition/deletion, the {\arrayring} representation is of course very powerful.
As in the case of {\arraylist}, the insertion anywhere other than head or tail remains problematic and requires potentially consequential moves.
Nevertheless, the {\arrayring} representation remains quite efficient when the insertion is close to either end ({\tt{}0} or {\tt{}size-1}).
Note that it is always better to have a capacity that is a power of 2.
In fact, the modulo that is necessary for the circular overflow of the indices is calculated using the bitwize operator {\tt{}and}\footnote{
Let $c$ be the capacity of the table which is a power of two.
Given a valid index $i$ in the table and an offset $\Delta$ with respect to that index,
the corresponding index is given by $((i\pm{}\Delta)${\tt{}\&}$(c-1))$.
If $c$ is statically known, the calculation will only take one processor cycle.
}.

The Fig. \ref{ArrayBlock} gives an example of the {\arrayblock} representation which is intended to behave more efficiently for all cases of insertions/deletions.
This representation consists in using a resizable primary table that allows access to secondary level tables, the blocks, which are all of the same size.
All blocks as well as the primary table itself are managed in a circular way, according to the same principle as for {\arrayring}.
The size of a secondary table is therefore a power of two, and relatively close to the size of memory page of the operating system, that is 2048 elements\footnote{
MMU (Memory Management Unit) is generally 4096 bytes in size.
This is equivalent to 512 words of 64 bits.
This choice of a 4 KB page was particularly well suited to 32-bit architectures.
However, it is generally accepted that the use of a larger table of 8 KB or even 16 KB is preferable on 64-bit architectures.
},
which is the value that gave the best performance.
Moreover, as in the case of lists, the representation {\arrayblock} has an index cache thanks to the variables {\tt{}cache\,prim} and {\tt{}cache\,index}.
The variable {\tt{}cache\,prim} is used to store the index of the block corresponding to the last access.
The variable {\tt{}cache\,index} returns the user index corresponding to the first data of the corresponding block.
Thus, as seen before, when memory accesses are located in a certain area,
ideally close to {\tt{}cache\,index}, we can restart the search from the block corresponding to the {\tt{}cache\,prim} index.
The strategy of the insertion and deletion algorithms is to preserve as much as possible, about a third, for free spaces within each block.
In this way we avoid shifts in the primary table as much as possible.
In this article, we will not go into detail about the insert and delete strategies, which can be very different and whose effectiveness depends mainly on the tests performed.

Without claiming to be completely exhaustive, these three array-based representations, {\arraylist}, {\arrayring} and {\arrayblock} provide a fairly complete overview.

\section{The Bjarne Stroustrup benchmark}

\begin{algorithm}
\caption{{\hfill}The Bjarne Stroustrup benchmark}\label{algostroustrup}
{\small\begin{algorithmic}[1]
\State $list\gets emptyList()$   
\For{$i\gets 1,N$}\Comment{Step 1: filling of $list$}
   \State $val\gets$ random number
   \State $idx\gets 0$
   \While{$(idx \leq size(list)-1) \wedge (item(list,idx) < val)$}
      \State $idx\gets idx + 1$
   \EndWhile
   \State $list\gets insert(list,idx,val)$
\EndFor
\For{$i\gets 1,N$}\Comment{Step 2: clearing of $list$}
   \State $idx\gets$ random in $[0,size(list)-1]$
   \State $list\gets remove(list,idx)$ ;
\EndFor
\end{algorithmic}}
\end{algorithm}

The benchmark proposed by B. Stroustrup consists of two phases (see algorithm \ref{algostroustrup}).
The first phase consists, for a given $N$ value, in progressively building a sorted list composed of $N$ randomly selected values.
The second phase consists in removing the $N$ values one by one, by randomly choosing the index of the removed value for each removal.
Note that during the first phase of the insertion, as indicated by B. Stroustrup, we naively and sequentially search for the right position to make the insertion.
It is not a dichotomous search for the right place to insert, as one might think.

Fig. \ref{stroustrup1} shows the results for the B. Stroustrup benchmark with all the data structures previously described.
Without having to go very far for the value of $N$,
as announced by B. Stroustrup, there is a clear separation between linked and array-based implementations.
In addition, this first run also shows the importance of having a cache for the last access index.
In fact, the {\nocachelist} implementation is very slow already for a very small value of $N$.
Even if it is possible to integrate the index memory cache into an iterator,
it is still preferable to integrate it directly into the list as soon as the manipulation interface allows access to the elements via an indexing mechanism.

\benchmarkfig{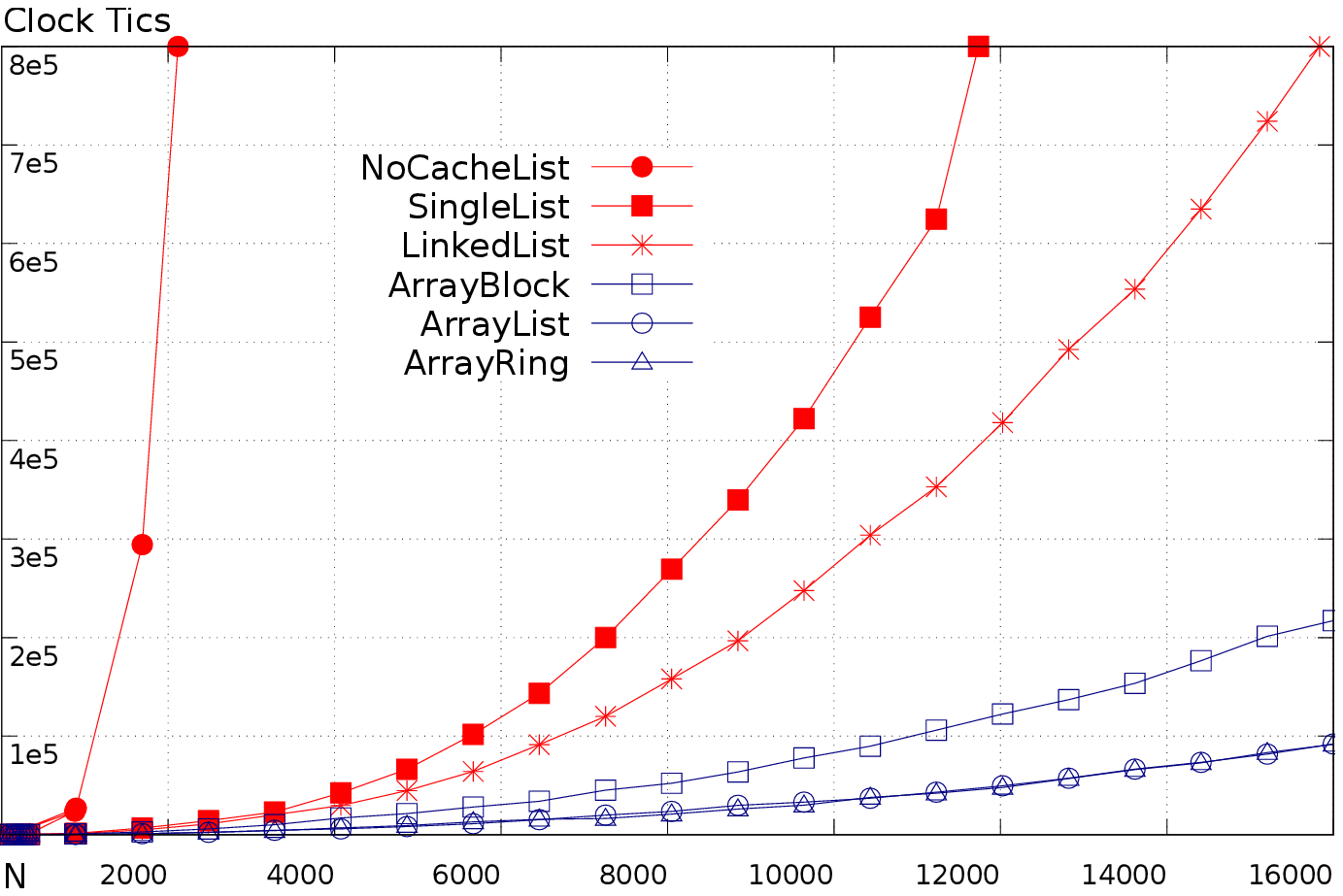}{
  The benchmark of B. Stroustrup using all the implementations of the paper, emphasizing limited $N$ values.
  Even for relatively small collections, a clear distinction is evident between linked lists (red) and array-based structures (blue).
}

Still on the Fig. \ref{stroustrup1} and still on chained implementations,
we can see the interest of the bidirectional linking, between {\singlelist} and {\linkedlist}.
In fact, thanks to double chaining, it is possible to go backwards from the index cache, which is not possible with single chaining.
As one might expect, {\singlelist} should be reserved for algorithms that essentially only traverse in the ascending direction of the indices (i.e. from left to right).

The three best results are obtained with array-based representations: {\arraylist}, {\arrayring}, and {\arrayblock}.
Note here that {\arrayblock} is significantly slower than the other two array-based representations.
Indeed, on arrays of relatively small size, as in this initial benchmark, using {\arrayblock} results in a time loss.

Note in passing that, in practice, it is essential for the programming language being used to facilitate a seamless switch between representations.
In object-oriented languages, when the library is well-designed, the change of representation is achieved by modifying only the collection creation instruction.
It is the mechanism of dynamic binding, or even better, the compiler that statically takes care of redirecting operations to the corresponding implementation.

Disregarding {\nocachelist}, Fig. \ref{stroustrup2} also illustrates the execution of B. Stroustrup's code, extending the $N$ value.
However, even though the three array-based implementations are clearly more efficient than the chaining-based ones,
the execution times deteriorate very quickly for values of $N$ that remain very modest.
The complexity induced by the sequential insertion algorithm during the first insertion phase is of the order of $O(N^2)$ in direct correlation with our experimental results.
Still referring to Fig. \ref{stroustrup2}, even if $N$ is greater, using {\arrayblock} still results in a non-negligible time loss (around 2 times slower).

\benchmarkfig{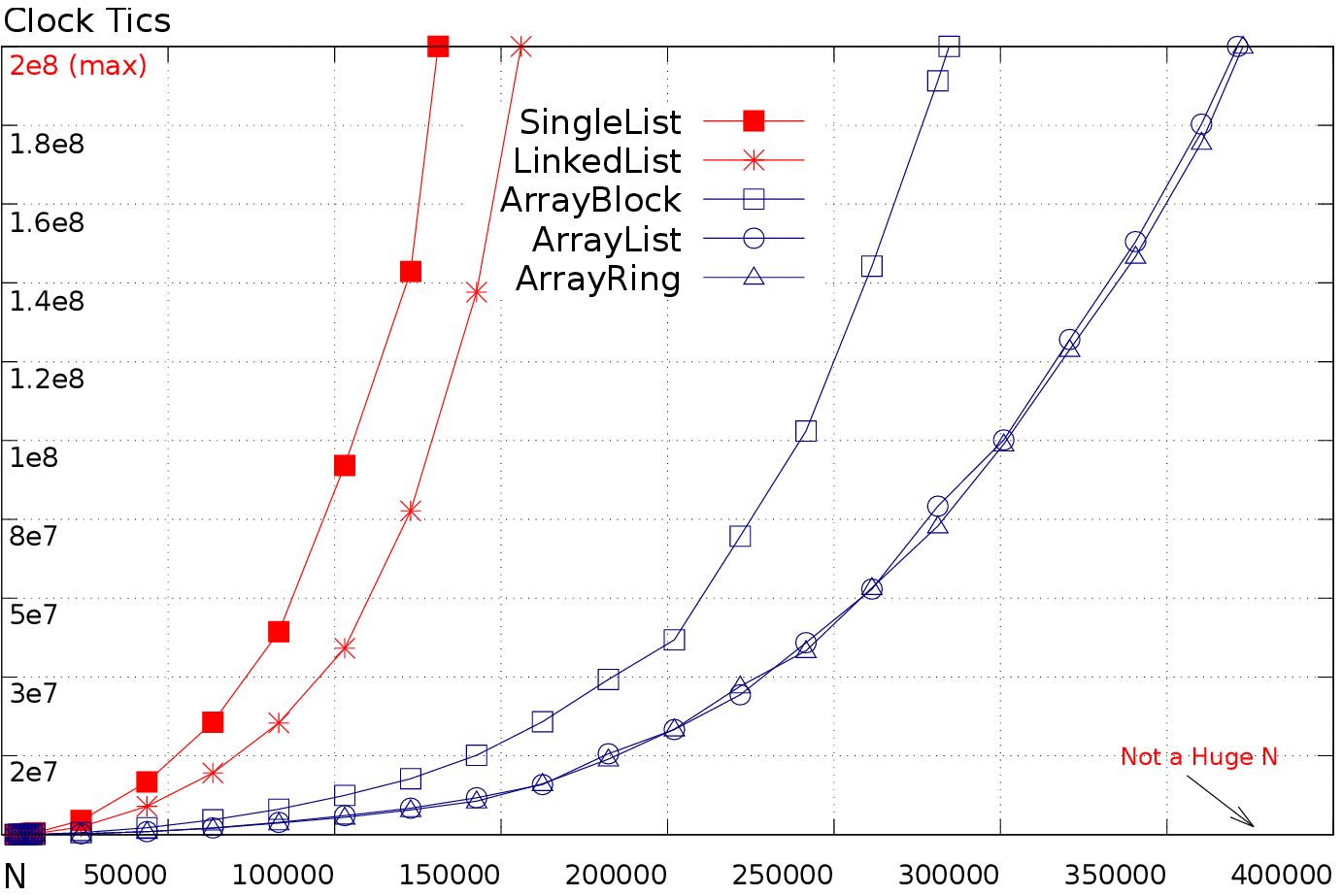}{
  With B. Stroustrup's benchmark, but pushing further for the value of N.
  The excessively slow representations are no longer shown (you will no longer see {\nocachelist}).
}

To visualize the relevance of the {\arrayblock} structure in the case of random insertion/deletion on large data structures,
we have slightly modified B. Stroustrup's benchmark by replacing the sequential search for the insertion location (lines $4$ to $7$ of the algorithm \ref{algostroustrup})
with a dichotomous search, which reduces the complexity of the first phase of the benchmark to $O(log_2(N))$.
The results for this modified version of the benchmark are shown in Fig. \ref{stroustrup3}.
The best results are clearly obtained with the {\arrayblock} implementation.
In fact, for the {\arraylist} and {\arrayring} implementations, a deletion or an insertion implies on average a shift of $N/2$ elements.
For {\arrayblock}, the number of elements to move does not depend on $N$; the shift are directly related to the constant size of a block.

\benchmarkfig{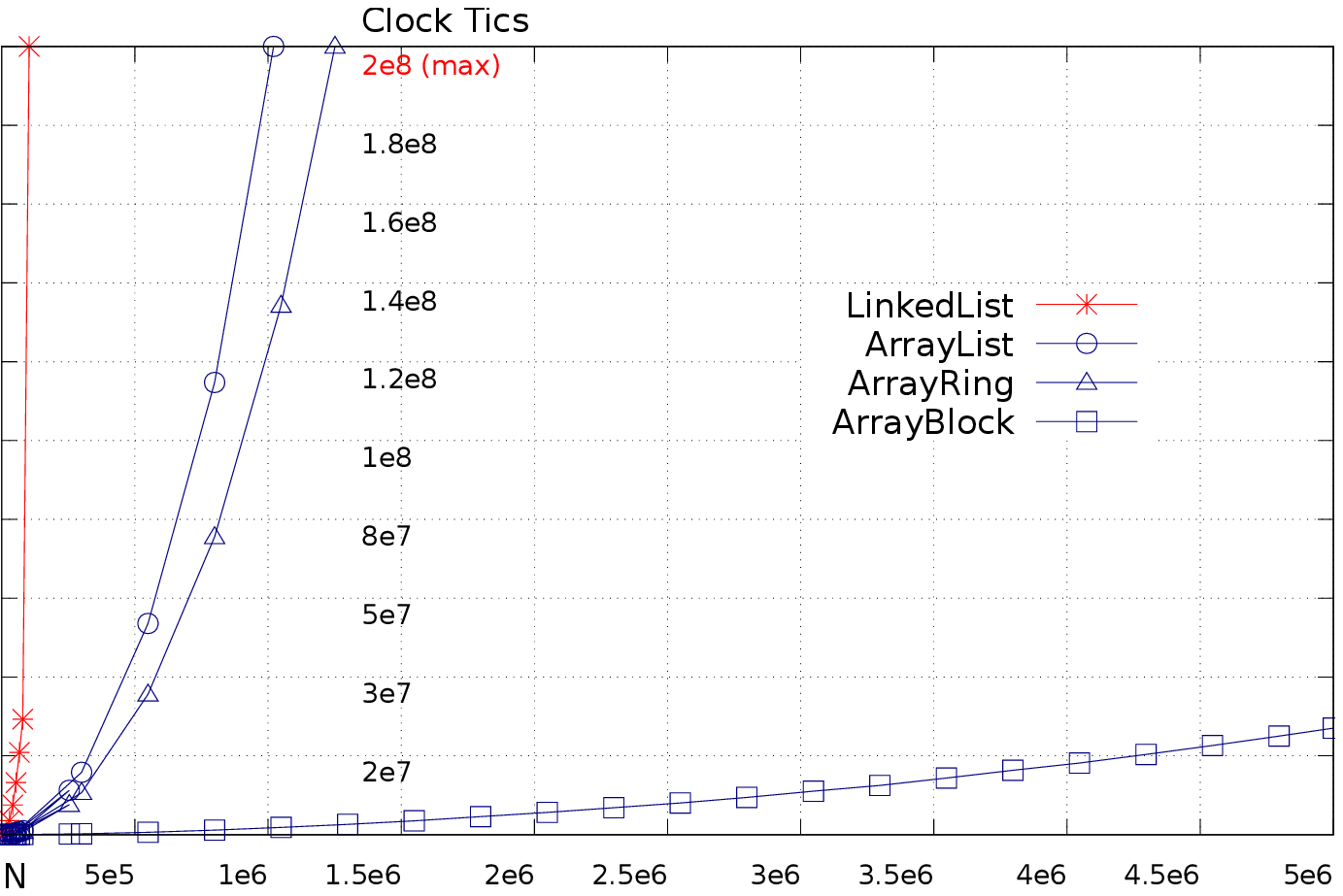}{
  Variation of benchmark B. Stroustrup: dichotomous insertion during the first phase.
  It is then possible to use a larger list by using {\arrayblock}.
  Memory saturation would take too much time.
}

Note that {\arrayring} performs marginally better than {\arraylist} because it allows choosing the most advantageous direction for shifting the elements.
As for the bad performance of {\linkedlist}, the problem does not come from the insertion or deletion which is in constant time,
but from the random access into the list which has an average complexity of $O(N/4)$.

\section{Fairbench: just fine for linked lists}

\begin{algorithm}
  \caption{{\hfill}The fairbench: the right benchmark for lists.}\label{algofairbench}
{\small\begin{algorithmic}[1]
\State $list\gets emptyList()$ ;   ~~ $idx\gets 0$ 
\For{$i\gets 1,N$} \Comment{Step 1: list filling}
   \If{$(i/N < 1/3)$}
      \State $list\gets addLast(list,someData(i))$ 
   \ElsIf{$(i/N < 2/3)$}
      \State $list\gets addFirst(list,someData(i))$ 
   \Else
      \State $idx\gets idx+1$ \Comment{or\,random\,incr.\,Fig.\,\ref{fairbench3}/\ref{fairbench4}/\ref{fairbench5}}
      \State $list\gets insert(list,idx,someData(i))$ 
   \EndIf
\EndFor
\For{$i\gets 1,N$} \Comment{Step 2: list traversal}
\State $sum\gets sum+value(list,i)$
\EndFor
\State $idx\gets N/2$ ; \Comment{Step 3: list clearing}
\For{$i\gets 1,N$}
   \If{$(i/N < 1/3)$}
      \State $idx\gets idx-1$ ; \Comment{or\,random\,decr.\,Fig.\,\ref{fairbench3}/\ref{fairbench4}/\ref{fairbench5}}
      \State $list\gets remove(list,idx)$ 
   \ElsIf{$(i/N < 2/3)$}
      \State $list\gets removeFirst(list)$
   \Else
      \State $list\gets removeLast(list)$
   \EndIf
\EndFor
\end{algorithmic}}
\end{algorithm}

The idea of the fairbench (see algorithm \ref{algofairbench}),
is to design a benchmark that is truly adapted to the concept of a list, in order to know whether linked implementations have good performance compared to array-based implementations,
and this for a consequent value of $N$.

The first phase (see {\it{}step 1} of the algorithm \ref{algofairbench}) for the fairbench is to grow the collection to its maximum of $N$
using only {\tt{}addLast} for the first third, then using only {\tt{}addFirst} for the second third,
and finally adding the last third during a single sequential run in the direction of increasing indices.
Before emptying the list completely, we perform a complete run from right to left to calculate the sum of all the values (see {\it{}step 2} of the algorithm \ref{algofairbench}).
The third and last phase consists in emptying the list by proceeding in the opposite way to the filling phase (see {\it{}step 3} of the algorithm \ref{algofairbench}).
This benchmark is clearly designed to benefit linked lists as much as possible.

\benchmarkfig{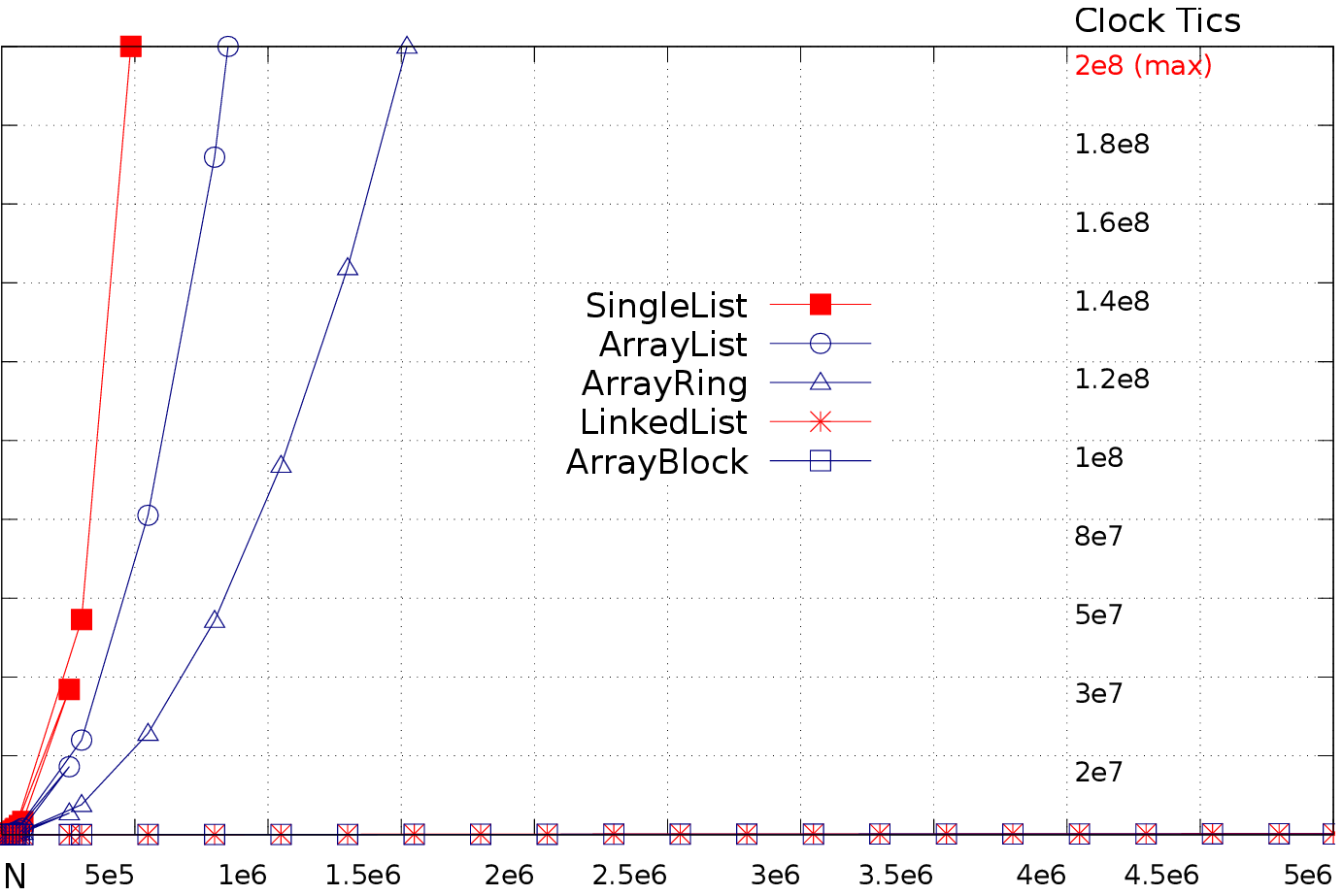}{
  Fairbench: to provide a chance for linked lists to win.
  Without going too far for $N$.
  And the winners are: {\linkedlist} and {\arrayblock}.
}

Fig. \ref{fairbench1} shows the results of fairbench without going too far for the value of $N$ in order to be able to distinguish which implementations are eliminated first.
This Fig. clearly separates the losers ({\singlelist}, {\arraylist} and {\arrayring}) from the winners ({\linkedlist} and {\arrayblock}).
Without being ridiculous, the value of $N$ separating the winners from the losers, remains relatively modest considering the memory of today's computers.
Note that the poor performance of {\singlelist}, a list with single linking, is mainly explained by the use of {\tt{}removeLast} in this benchmark.

\benchmarkfig{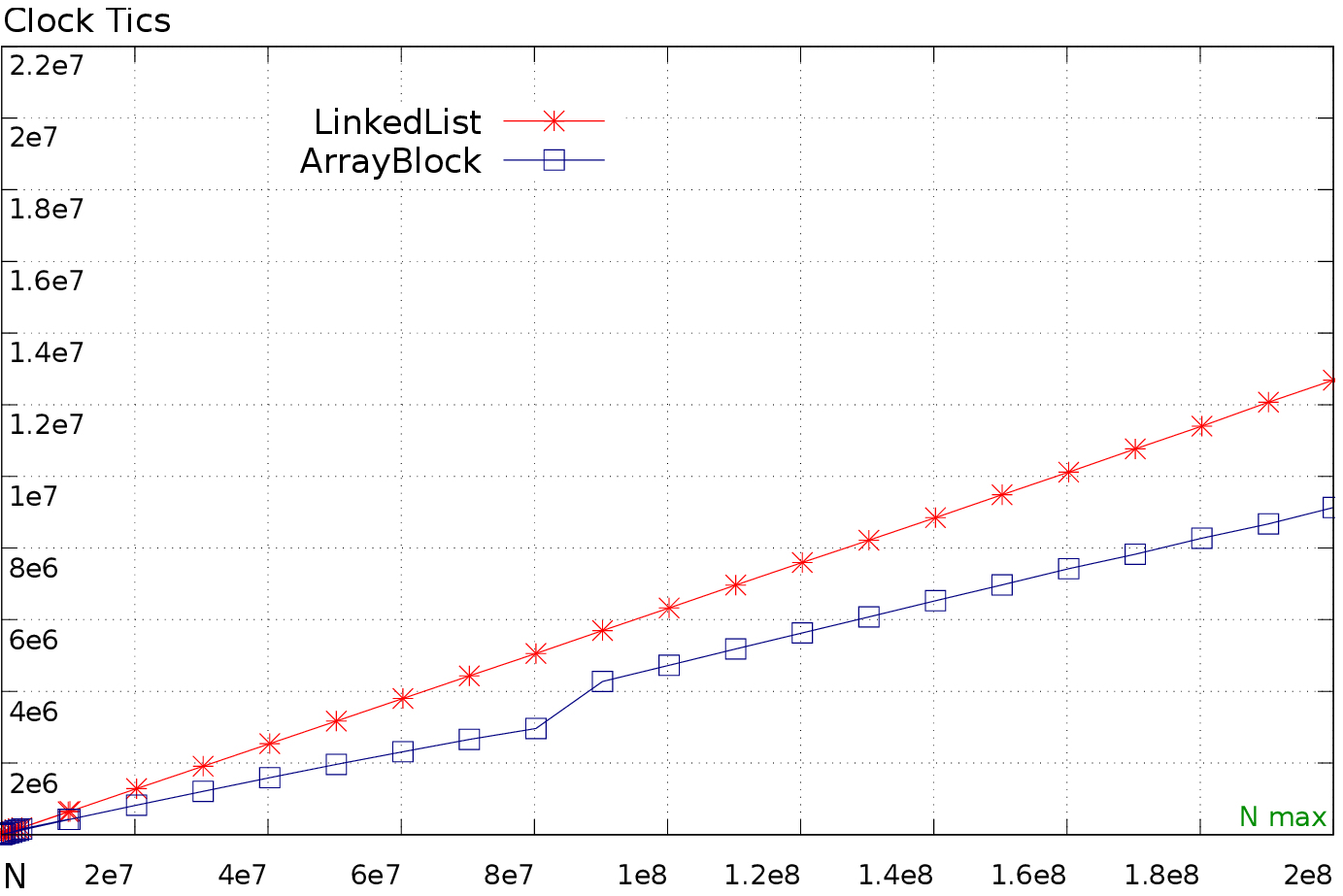}{
  Fairbench with $N$ that saturates a 16 GB (gigabyte) memory in order to determine the true winner.
  {\linkedlist} needs around 13 GB where {\arrayblock} needs around 10 GB for the largest value of $N$ (see Section \ref{benchmarking} for details).
}

Fig. \ref{fairbench2} shows the results of running fairbench maximizing the value of $N$ over the memory of the computer used for
testing\footnote{
All the subsequent figures are designed to maximize the value of $N$, as denoted by the green tag located in the bottom right corner.
}.
Of all the data structures presented, {\linkedlist} is the most memory hungry, so it is {\linkedlist} that determines the maximum possible value of $N$
(see Section \ref{benchmarking} for more details on the computers/compilers used).

Although fairbench is designed to favor linked implementations, it is still {\arrayblock} which behaves better than {\linkedlist}.
However, the difference in execution speed, while already noticeable, did not correspond to our observations during
development\footnote{
We use {\arrayblock} in the development of large-scale software, including compilers and robotics systems,
and noticed an even more pronounced difference when transitioning from {\linkedlist} to {\arrayblock}.
}.
Upon closer examination of the excellent results achieved by {\linkedlist}, it becomes evident that fairbench tends to optimize for the precise memory caching of {\linkedlist}.

Indeed, the memory cache of a {\linkedlist} refers to the memory location of a single element (see Fig. \ref{LinkedList}).
If access to the element immediately before or after is necessary, the memory cache of the {\linkedlist} must be updated accordingly.
Regarding {\arrayblock} (see Fig. \ref{ArrayBlock}), the memory cache designates an entire block, requiring fewer updates.
As long as two consecutive accesses do not vary more than the average number of elements per block, the cache of an {\arrayblock} often remains valid.
Thus, in Algorithm \ref{algofairbench}, we attempted to replace the index variations in lines 8 and 18 with random variations.

\benchmarkfig{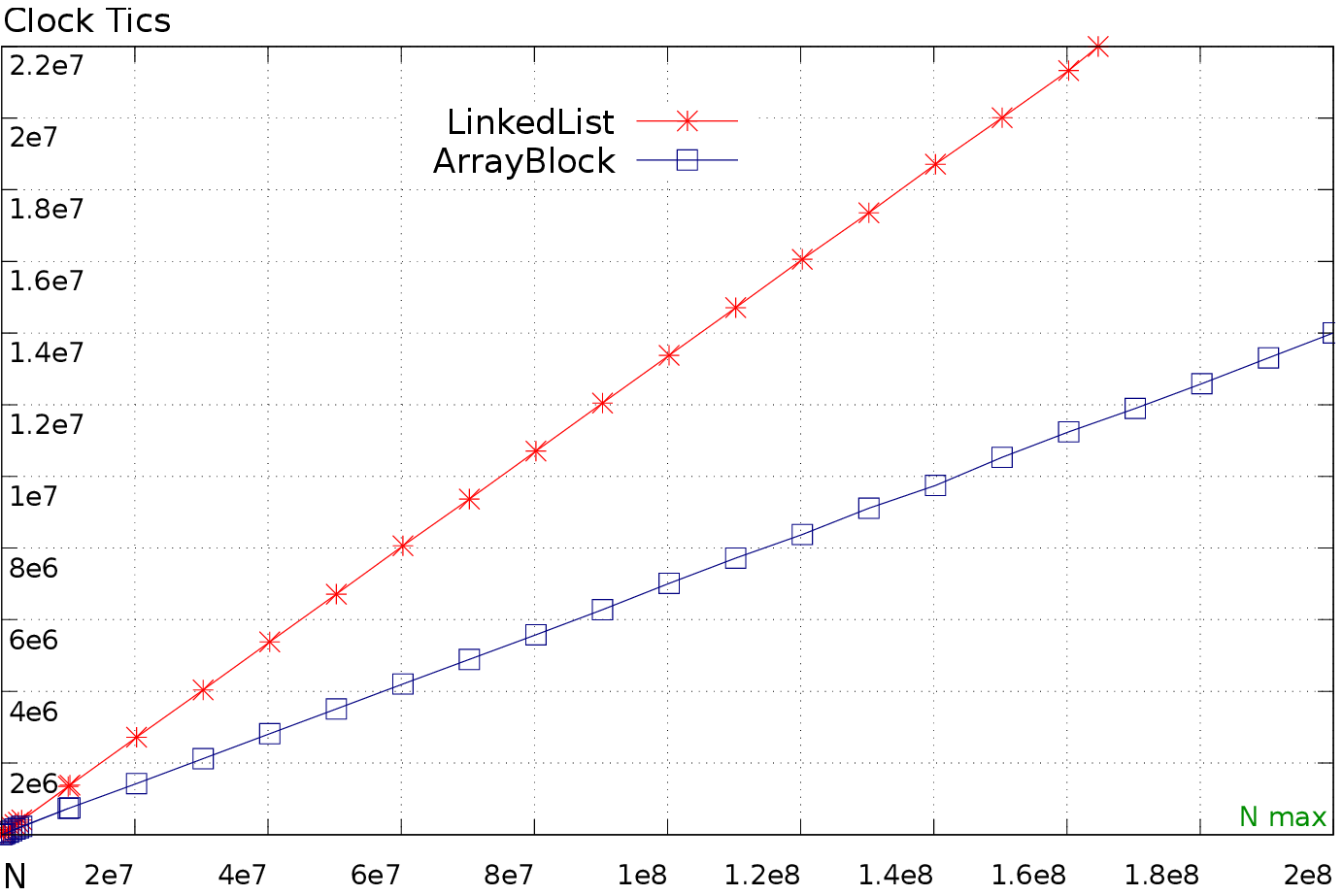}{
  Fairbench with modifications of lines 8 and 18 in the algorithm \ref{algofairbench}.
  The variable $idx$ is incremented/decremented using a random number in range $[1,32]$.
}

\benchmarkfig{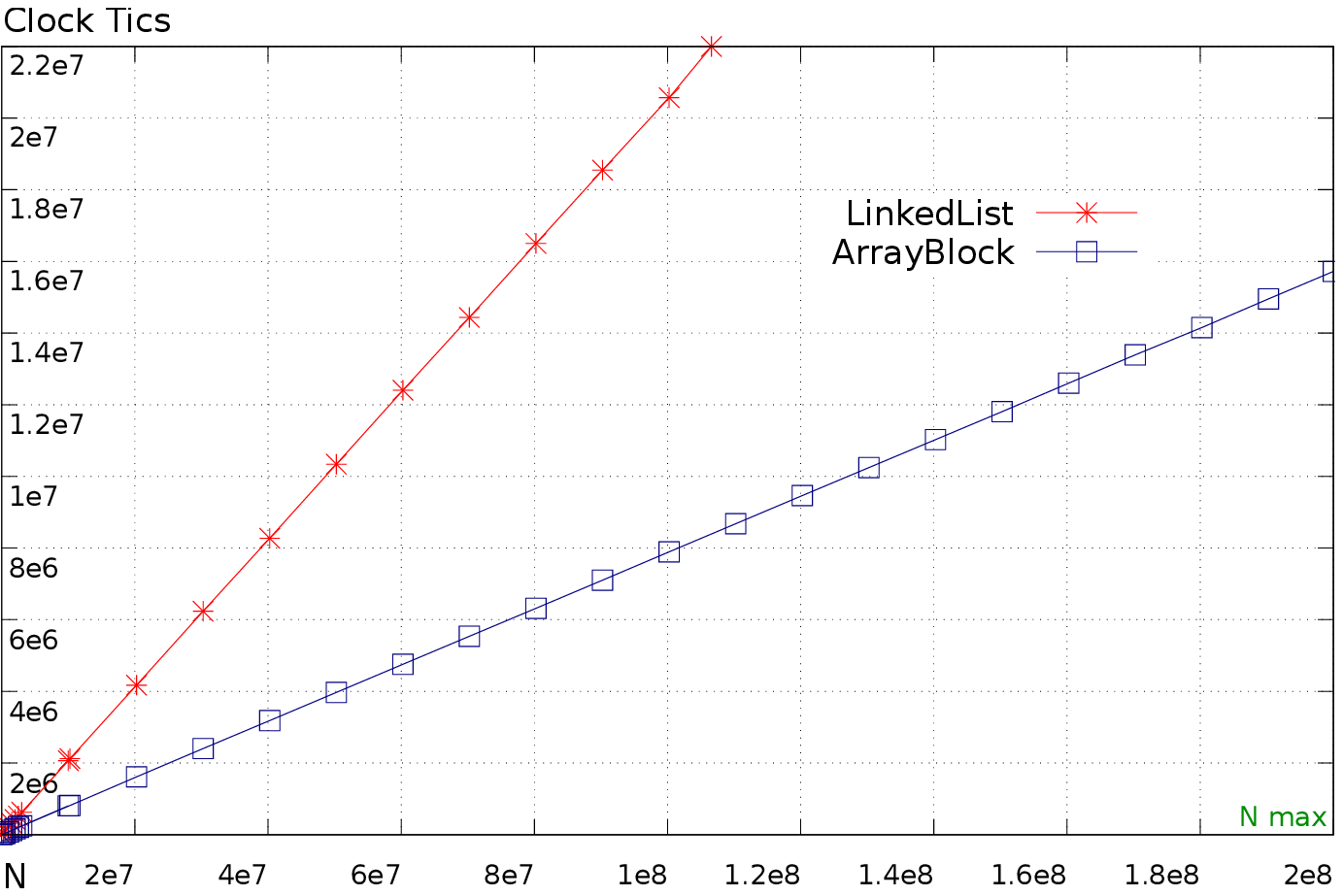}{
  Fairbench with modifications of lines 8 and 18 in the algorithm \ref{algofairbench}.
  The variable $idx$ is incremented/decremented using a random number in range $[1,64]$.
}

\benchmarkfig{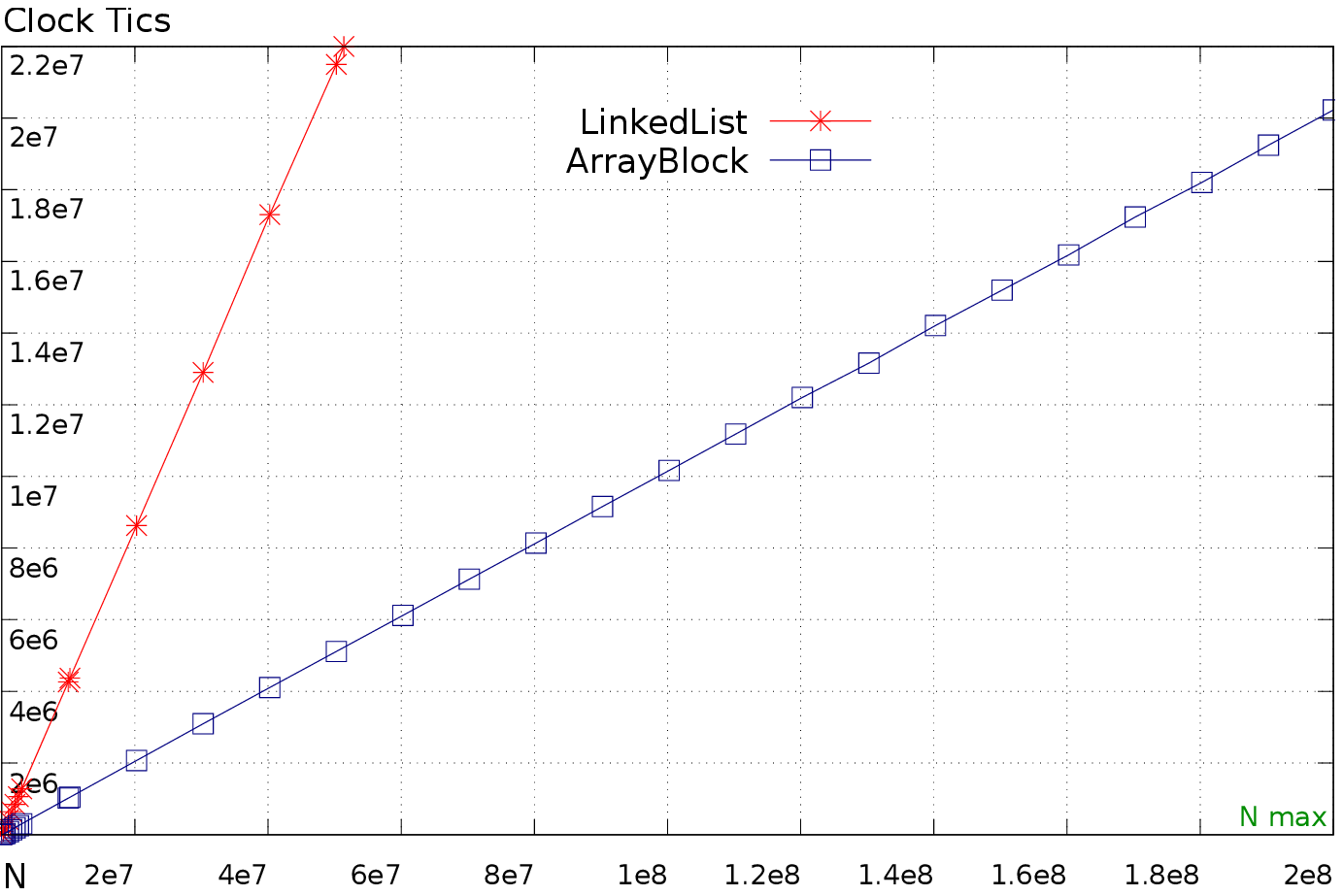}{
  Fairbench with modifications of lines 8 and 18 in the algorithm \ref{algofairbench}.
  The variable $idx$ is incremented/decremented using a random number in range $[1,128]$.
}

After several attempts, the hypothesis of a memory cache too specific for linked lists seems to be proving correct.
Thus, in Figures \ref{fairbench3}, \ref{fairbench4}, and \ref{fairbench5}, we replace the increment/decrement of 1 with a progressively larger random increment/decrement.
As we can see, the situation deteriorates significantly with the increasing increments.
That being said, it is evident that these modifications to the Fairbench algorithm are, in a way, a return to Bjarne Stroustrup's benchmark.
Nevertheless, it is observed that on collections of significant sizes, it is always {\arrayblock} that achieves the best performance.

\section{Endgame: Add First or Remove Last}

\begin{algorithm}
\caption{~{\hfill}addLast / traversal / removeLast benchmark}\label{addlastalgo}
{\small\begin{algorithmic}[1]
\State $list\gets emptyList()$   
\For{$i\gets 1,N$}\Comment{Step 1: filling of $list$}
   \State $list\gets addLast(list,randomNumber)$
\EndFor
\For{$i\gets 1,N$} \Comment{Step 2: list traversal}
\State $sum\gets sum+value(list,i)$
\EndFor
\For{$i\gets 1,N$}\Comment{Step 3: clearing of $list$}
   \State $list\gets removeLast(list)$ ;
\EndFor
\end{algorithmic}}
\end{algorithm}

\benchmarkfig{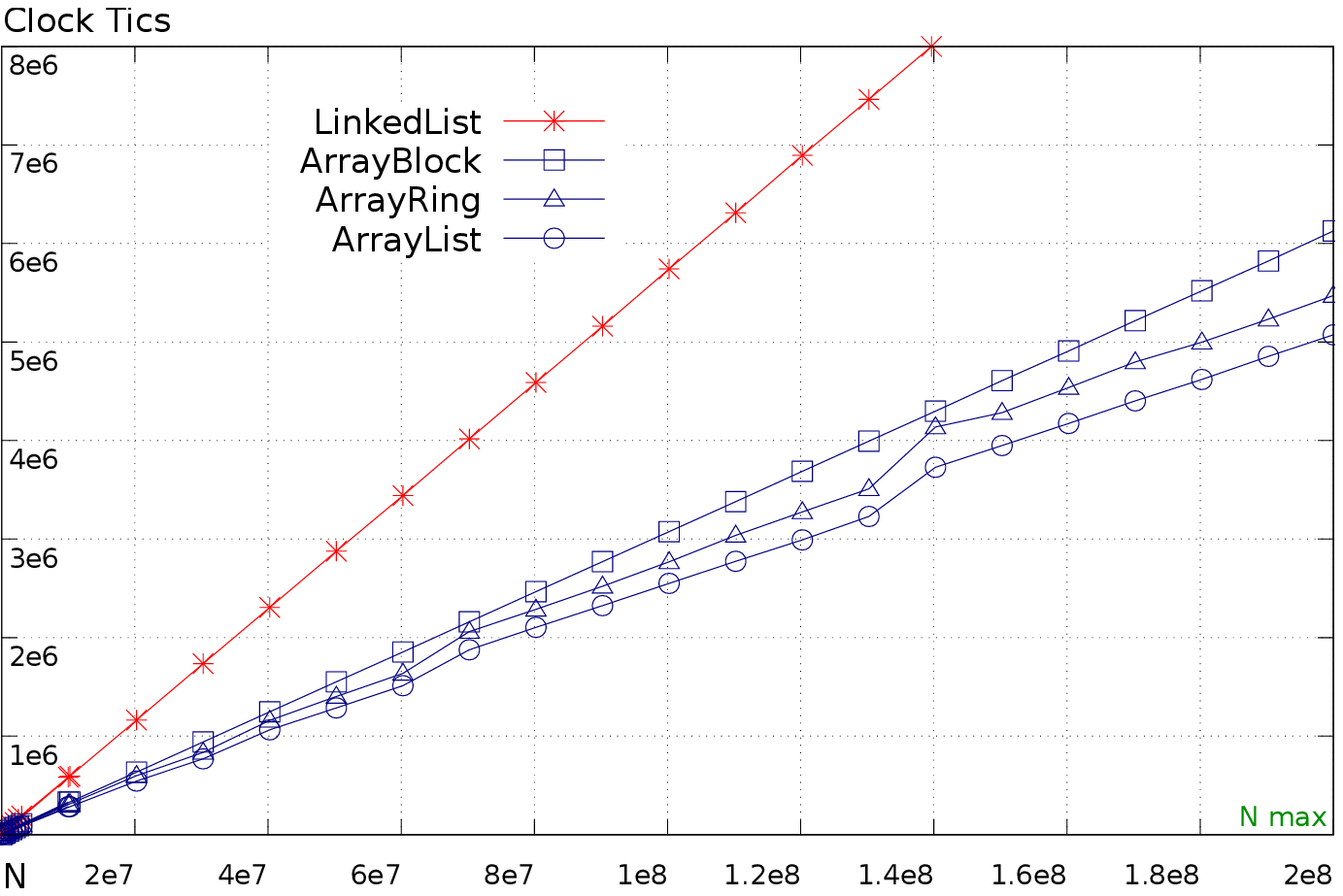}{
  addLast / traversal / removeLast (algorithm \ref{addlastalgo}).
}

To complete our comparison, we have added two benchmarks, both of which also clearly favor linked representations.
One of them favors adding and deleting at the head of the list: we add only with {\tt{}addFirst}, and we empty the list only with {\tt{}removeFirst}.
The other one favors adding and deleting in queue: we add only with {\tt{}addLast}, and we empty the list only with {\tt{}removeLast}.
In both cases, before emptying the list, we make a single run from left to right, thus accessing all the elements.

Moreover, since we are only interested in lists of a significant size, we saturate 16 GB of memory in both cases and keep only the implementations that withstand these constraints.
The results are shown in figures \ref{addlast} and \ref{addfirst}.
Even though these results speak for themselves, we should mention that it is not possible to add {\singlelist} in Fig. \ref{addlast} because in the case of {\tt{}removeLast},
the complexity is about $O(n)$.
No surprise, it is also not possible to present {\arraylist} on the Fig. \ref{addfirst}.

\begin{algorithm}
\caption{{\hfill}addFirst / traversal / removeFirst benchmark}\label{addfirstalgo}
{\small\begin{algorithmic}[1]
\State $list\gets emptyList()$   
\For{$i\gets 1,N$}\Comment{Step 1: filling of $list$}
   \State $list\gets addFirst(list,randomNumber)$
\EndFor
\For{$i\gets 1,N$} \Comment{Step 2: list traversal}
\State $sum\gets sum+value(list,i)$
\EndFor
\For{$i\gets 1,N$}\Comment{Step 3: clearing of $list$}
   \State $list\gets removeFirst(list)$ ;
\EndFor
\end{algorithmic}}
\end{algorithm}

\benchmarkfig{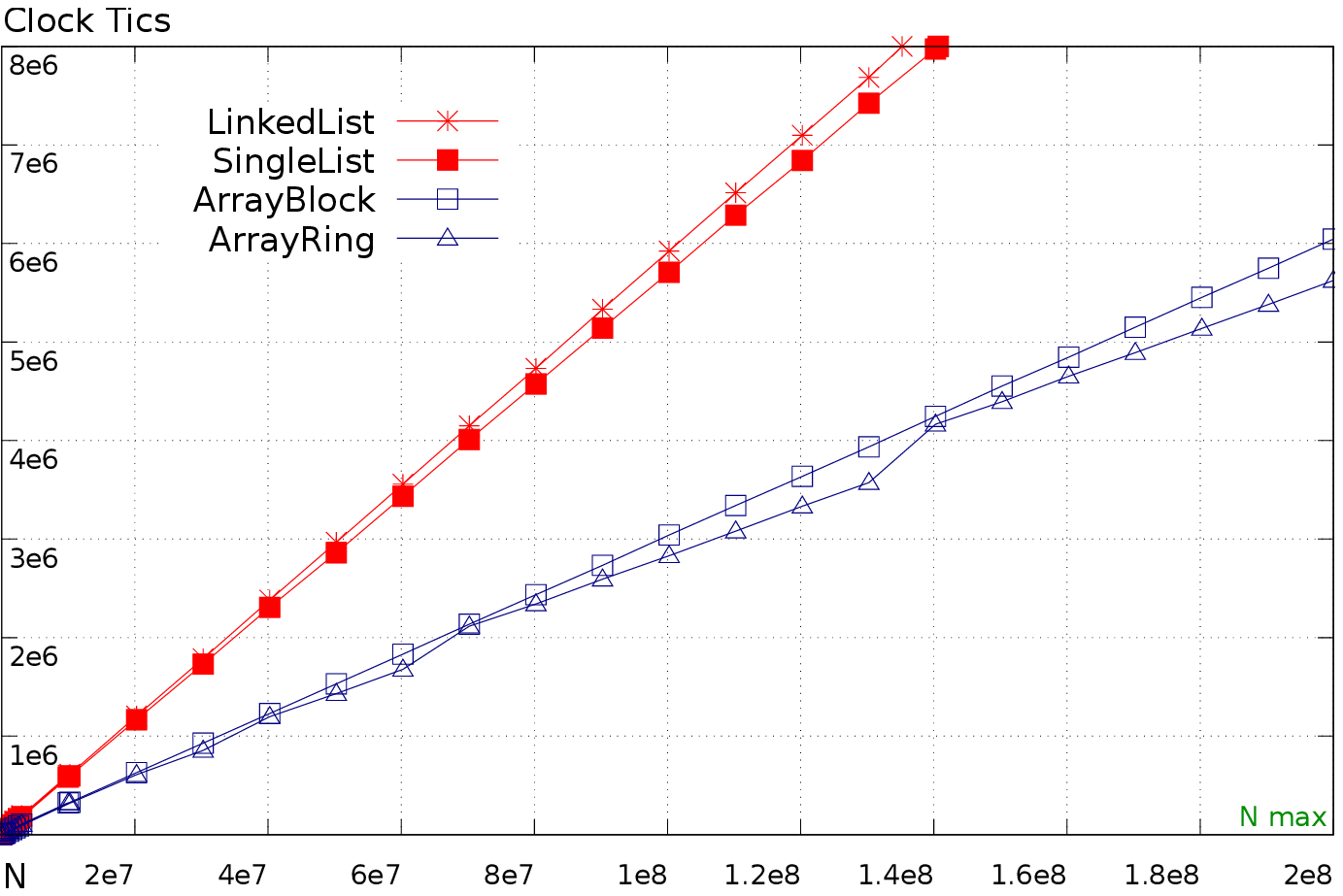}{
  addFirst / traversal / removeFirst (algorithm \ref{addfirstalgo}).
}

\section{Benchmarking Methodology}\label{benchmarking}
All the benchmarks presented in this article were conducted on the same computer configuration, consistently utilizing the same programming language,
the same compiler with identical options, and the same library.

We have, of course, verified that the change of processor has no effect.
Whether using an Intel Core or AMD processor is inconsequential.
No relative differences were observed following the processor change. The performance curves exhibit a remarkable relative stability.

The tests were conducted on memory capacities ranging from 4 to 16 GB, and similarly, no relative disparities were detected across these different configurations.
The curves presented in the article correspond to a 16 GB memory, and the maximum value of $N$ ($N_{\text{max}} = 2 \times 10^8$)
was chosen to reach this maximum value with the most memory-intensive data structure (see Fig. \ref{fairbench2}).
For those of you who will rerun the tests on your own machine, don't forget to decrease or increase the maximum value of $N$ based on the maximum size of your
memory\footnote{A constant is provided for this purpose in the code accompanying the article.}.

Furthermore, it should be noted that transitioning from one operating system to another, whether it be Linux or MacOS, had no notable impact.
It is highly conceivable that the results remain consistent when using Microsoft Windows or, of course, also an Android smartphone.

Regarding execution times, we chose to display only the number of clock ticks, considering that an absolute value in minutes or seconds would not add any significant
information\footnote{Nevertheless, it is worth noting that it took approximately 5 hours to recalculate all the figures in this article.}.
Additionally, each execution is repeated multiple times, and only the minimal value in terms of execution time is retained.
Working on a machine dedicated to benchmark execution, we also observed a rapid convergence towards a stable result.
Thus, after 3 or 4 executions of the same test, we converge towards a minimal value that remains stable thereafter.

It is during the implementation of a project combining compilation and robotics that we developed our own data structure, named {\arrayblock}.
The outstanding results achieved with {\arrayblock} during these developments motivated the writing of this article.

Regarding the language utilized in the context of our robotics project, it is important to note that we used the {\lisaac} programming language \cite{sonntag2002a} \cite{sonntag2013}.
This language allows us to achieve optimal performance, and we primarily use it because it generates excellent C code.
The interfacing with robotics peripherals is thus straightforward.
Since {\lisaac} code is initially translated into C before benefiting from all the optimizations of the C compiler,
it is evident that we would gain nothing by writing all these benchmarks directly in C.

Furthermore, since the {\lisaac} language is an object-oriented language, all benchmark algorithms are written only once and shared through inheritance.
The {\collection} type serves as a common abstraction for all the data structures presented in this article.
Therefore, each time, the same code inherited from {\collection} is applied to all the different data structures being compared.

Regarding the compilation of the C code, all the benchmarks in this article were compiled with gcc.
The use or non-use of gcc optimization options has no relative effect.
The change of the C compiler also does not disturb the ranking.

\section{Conclusion}

Indeed, our study shows that it is very difficult to find much interest in using linked lists in real applications.
As long as we are on small values of $N$, linked implementations rarely have a significant advantage.
However, if we stick to the idea of using linked lists, this article also highlights the importance of implementing an index cache for the managing of a linked
list\footnote{
At the time we are writing this article, the \cpp{} {\tt{}std::list} data structure have no index memory cache.
}.
As mentioned earlier, the optimal approach is to integrate a cache within the data structure itself.
Additionally, if iterators also exist in your library, it is advisable to have a dedicated cache for each iterator \cite{colnet99}.

Overall, the increase in RAM size and the speed of today's processors allows more complex implementations to be used.
Also, the processor's memory caches give advantage to contiguous accesses and speed up data shifting.
Our {\arrayblock} implementation takes advantage of these technological innovations and gives very good results in most all the tests we have done.
In fact, as soon as the size of the collection becomes substantial, {\arrayblock} is likely the best choice.
However, we could invent and test many types of benchmarks, but we focused on extreme cases that theoretically give the linked list an advantage.

The memory representation of {\arrayblock} is very similar to the more basic implementation of the virtual memory management on current processors.
The two levels of the MMU indirection table on 32-bit processors (4 levels for 64-bit processors) are similar to our primary table.
Then, the fixed 4KB pages are similar to our small circular arrays of fixed size in powers of 2.
The primary table gives the flexibility to add non-contiguous blocks for fast insertions, and our small contiguous arrays bring the speed of direct access to an element.
The circular index management for both the primary table and the small contiguous arrays allows the complexity of adding or deleting to be divided by two.
The cost of a circular index management is negligible compared to the benefits.
We show it perfectly here with quite surprising results with a simple implementation of the circular management of an array {\arrayring}.

Note that currently, the implementation of
Python\footnote{CPython 3.12.1 file \texttt{Objects/listobject.c}}
uses a data structure equivalent to an {\arraylist} for the native $[]$ operator, even in the case of a substantial collection.
For high-level languages designers, the search for an ideal list structure implementation under all circumstances is also important.
A high-level language whose goal is to simplify the choice of data structures by using a single structure for lists,
especially for untyped languages,
must pay attention to this implementation or else its overall performance will be severely degraded.
For this important issue, our implementation is clearly a very polyvalent solution.

\newpage % DCDC ajustement des 2 colonnes en fin d'article

\section*{Conflict of Interest}
The authors declare no conflict of interest.

\section*{Author Contribution}
The authors contributed equally to this work and are listed in alphabetical order of their first names.

\section*{Funding}
Both authors are civil servants of the French government.
We gratefully acknowledge the French government for providing us with the freedom to pursue our research interests.

%\IEEEtriggeratref{7}

\bibliographystyle{plain}
\bibliography{biblio}
%\tableofcontents
\end{document}